\documentclass[%
 aip,
 jmp,%
 amsmath,amssymb,
preprint,%
 reprint,%
]{revtex4-1}

\usepackage{graphicx}
\usepackage{dcolumn}
\usepackage{bm}

\begin{document}

\preprint{AIP/123-QED}

\title{ YBCO microwave resonators for strong collective coupling with spin ensembles  }

\author{A. Ghirri}
\email{alberto.ghirri@nano.cnr.it.} 
\affiliation{Istituto Nanoscienze - CNR, Centro S3, via Campi 213/a, 41125 Modena, Italy}

\author{C. Bonizzoni}
\affiliation{Dipartimento Fisica, Informatica e Matematica, Universit\`{a} di Modena e Reggio Emilia and Istituto Nanoscienze - CNR, Centro S3, via Campi 213/a, 41125 Modena, Italy.}

\author{D. Gerace}
\affiliation{Dipartimento di Fisica, Universit\`{a} di Pavia, via Bassi 6, 27100 Pavia, Italy.}

\author{S. Sanna}
\affiliation{Dipartimento di Fisica, Universit\`{a} di Pavia, via Bassi 6, 27100 Pavia, Italy.}

\author{A. Cassinese} 
\affiliation{CNR-SPIN and Dipartimento di Fisica, Universit\`{a} di Napoli Federico II, 80138 Napoli, Italy.}

\author{M. Affronte}
\affiliation{Dipartimento Fisica, Informatica e Matematica, Universit\`{a} di Modena e Reggio Emilia and Istituto Nanoscienze - CNR, Centro S3, via Campi 213/a, 41125 Modena, Italy.}

\date{23 February 2015}

\begin{abstract}
Coplanar microwave resonators made of 330 nm-thick superconducting YBCO have been realized and characterized in a wide temperature ($T$, 2-100 K) and magnetic field ($B$, 0-7 T) range. The quality factor $Q_L$  exceeds 10$^4$ below 55 K and it slightly decreases for increasing fields, remaining 90$\%$ of $Q_L(B=0)$ for $B=7$ T and $T=2$ K. These features allow the coherent coupling of resonant photons with a spin ensemble at finite temperature and magnetic field. To demonstrate this, collective strong coupling was achieved by using DPPH organic radical placed at the magnetic antinode of the fundamental mode: the in-plane magnetic field is used to tune the spin frequency gap splitting across the single-mode cavity resonance at 7.75 GHz, where clear anticrossings are observed with a splitting as large as $\sim 82$ MHz at $T=2$ K. The spin-cavity collective coupling rate is shown to scale as the square root of the number of active spins in the ensemble.
\end{abstract}


\maketitle

Thanks to pioneering experiments and theoretical proposals, quantum technologies have enormously advanced and the interest can now be turned to explore viable routes for practical applications. Several pure quantum systems, including cold atoms, photons, superconducting qubits, spin impurities in Si, or nitrogen vacancies (NV) in diamond - among many others - have been deeply investigated in the last decade as potential candidate qubits for applications in quantum information processing, and techniques for their read out and manipulation have been developed.\cite{RevMPnori} Advantages and limitations of each system have been debated: whilst large margins of improvement are still possible for the different techniques, fundamental limits are clear for each system. A possible strategy to overcome these barriers is to combine quantum systems of different nature and take advantage of the best features of each of them in hybrid quantum devices. Of course, this opens new technological challenges. Along these lines, high-quality factor resonators play a pivotal role, since photons can be coupled with a number of other two-level systems (qubits) while begin optimal flying quantum bits themselves. Among them, planar resonators are particularly suitable to be coupled with a variety of atomic or solid state qubits, with the final goal of developing an on-chip hybrid quantum technology.\cite{RevMPnori, Grezes} In fact, mm-length microwave resonators can be routinely fabricated in a scalable arrangement and on different substrates. In particular, state-of-art superconducting resonators allow the achievement of power-independent quality factors as high as $10^6$ or above in planar geometry at the single photon level.\cite{Megrant, Minev} 

\begin{figure}[t]
\begin{center}
\includegraphics[width=8.5cm]{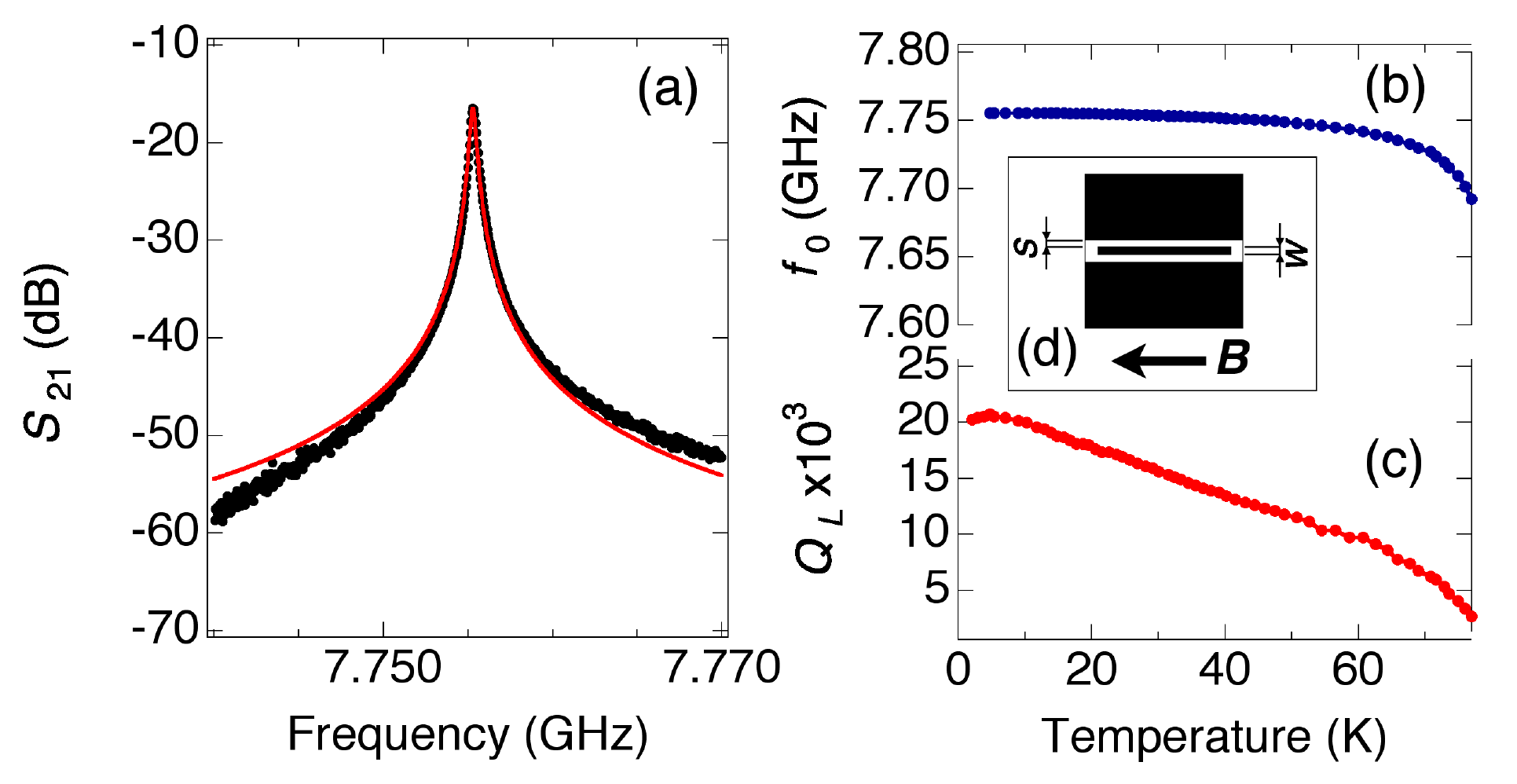}
\end{center}
\caption{Characteristics of bare YBCO resonators. (a) Transmission spectrum measured at $T=2$ K and $B=0$. The input power is $P_{inc}=-22.5$ dBm. The red line shows the calculated curve. Temperature dependence of the (b) resonance frequency, $f_0$, and (c) loaded quality factor, $Q_L$. In the inset (d) a sketch of the YBCO coplanar resonator is shown.}
\label{Fig1}
\end{figure}

A key step to coherently transfer information between cavity photons and stationary qubits is to achieve the strong coupling regime: electric or magnetic dipole coupling between the qubit and the confined electromagnetic field should overcome their respective damping rates.\cite{QOptics_Book} Electric dipole coupling allowed the observation of the strong coupling with single quantum emitters.\cite{Rempe,Brune,Wallraff,Hennessy} On the other hand, spin ensembles collectively coupled to microwave resonators have been proposed for a hybrid quantum technology.\cite{Imamoglu} Since the magnetic dipole coupling of a single spin to a resonator mode, $g_s$, is typically too small, a collective enhancement of the effective coupling rate of $N$ spins \cite{Dicke}, scaling as $g_s\sqrt{N}$, allows to overcome the limitations due to both the decoherence rate of the spin system, $\gamma_s$ and the inverse photon lifetime in the cavity, $\kappa=\omega_0/Q$ (where $f_0=\omega_0 / 2\pi$ is the resonant frequency and $Q$ is the resonator quality factor).\cite{TavisCummings} In this way, collective strong coupling of spin ensembles and microwave photons has been experimentally shown in coplanar resonators,\cite{Kubo,Amsuss,Schuster}  three-dimensional (3D) cavities,\cite{Chiorescu,Abe,Probst,Eddins} and microwave oscillators.\cite{Boero}
While 3D resonators are less suited for on-chip integration, all of the previous achievements employing planar resonators were obtained with conventional superconductors (typically Nb), which are limited to operate at moderate magnetic fields. However, manipulation of spins may need application of finite magnetic fields.\cite{Tabuchi, Zhang} Microwave resonators made of high $T_c$ superconductors, such as YBCO, have shown excellent performances from liquid Nitrogen temperatures \cite{Lancaster, Hein, Ghigo, YBCONapoli} down to mK range and single-photon regime.\cite{Arzeo} Thanks to the large value of their intrinsic upper critical field, these systems offer unprecedented possibilities for spin manipulation.

In this Letter, we show that YBCO coplanar resonators have excellent performances under strong magnetic fields, with quality factors significantly exceeding $10^4$ up to $T\sim 55$ K. Therefore, they appear as a significant step ahead for quantum technology applications. Stimulated by recent theoretical \cite{Chiorescu, Baibekov} and experimental \cite{Chiorescu, Abe, Boero} results, here we focused on the high photon number and high temperature regimes, where we report the strong collective coupling of an electron spin ensemble to YBCO microwave coplanar resonators. Our major interest is to use molecular spins, which offer several advantages with respect to spin impurities.\cite{GhirroTroianiAff2014} Interesting and sufficiently long phase memory times have been reported for simple radicals,\cite{Takui, Sorace} mono-metallic Cu phthalocyanine molecules,\cite{Aeppli} or (PPh$_4$)$_2$[Cu(mnt)$_2$] derivatives.\cite{CuMNT} 
In the present work, we employed commercial di(phenyl)-(2,4,6-trinitrophenyl)iminoazanium (DPPH), which is regularly used as field calibration marker in EPR spectroscopy. The decoherence time of DPPH is $T_1=T_2=62$ ns,\cite{EatonEaton} while the continuous wave linewidth is sharp ($\gamma_s / 2\pi\simeq 3.9$ MHz)\cite{Zilic} due to the exchange narrowing effect. Below 10 K the linewidth increases as an effect of antiferromagnetic interactions [$\gamma_s /2\pi\simeq 14$ MHz at 2 K].\cite{SovPhysJEPT} The strong coupling between DPPH radicals and a confined microwave field has been already demonstrated with 3D cavities\cite{Chiorescu, Abe} or microwave oscillators.\cite{Boero} 

Superconducting resonators were fabricated by optical lithography upon wet etching (2\% H$_3$PO$_4$ solution) of commercial 10 x 10 mm$^2$ double sided YBa$_2$Cu$_3$O$_7$ (YBCO, in short) films (330 nm thick) on sapphire  (430 $\mu$m) substrates (Ceraco Gmbh). The film is gold-coated on the back side to improve the contact to ground. The patterned coplanar structure is constituted by a 8 mm central strip having width $w=200$ $\mu$m and separation $s=73$ $\mu$m from the lateral ground planes [Fig. \ref{Fig1} (d)].\cite{Lancaster} The coupling of the resonator to the feed line can be adjusted by finely tuning the position of the launchers. We tested five YBCO planar resonators finding quite reproducible results.\cite{SI} Low temperature measurements were performed using a Quantum Design PPMS cryo-magnetic system equipped with 7 T magnetic field applied parallel to the plane of the resonator. Reflection [$S_{11}(f)$] and transmission [$S_{21}(f)$] scattering parameters were measured by means of an Agilent PNA Vector Network Analyzer (VNA). 

\begin{figure}[t]
\begin{center}
\includegraphics[width=8.5cm]{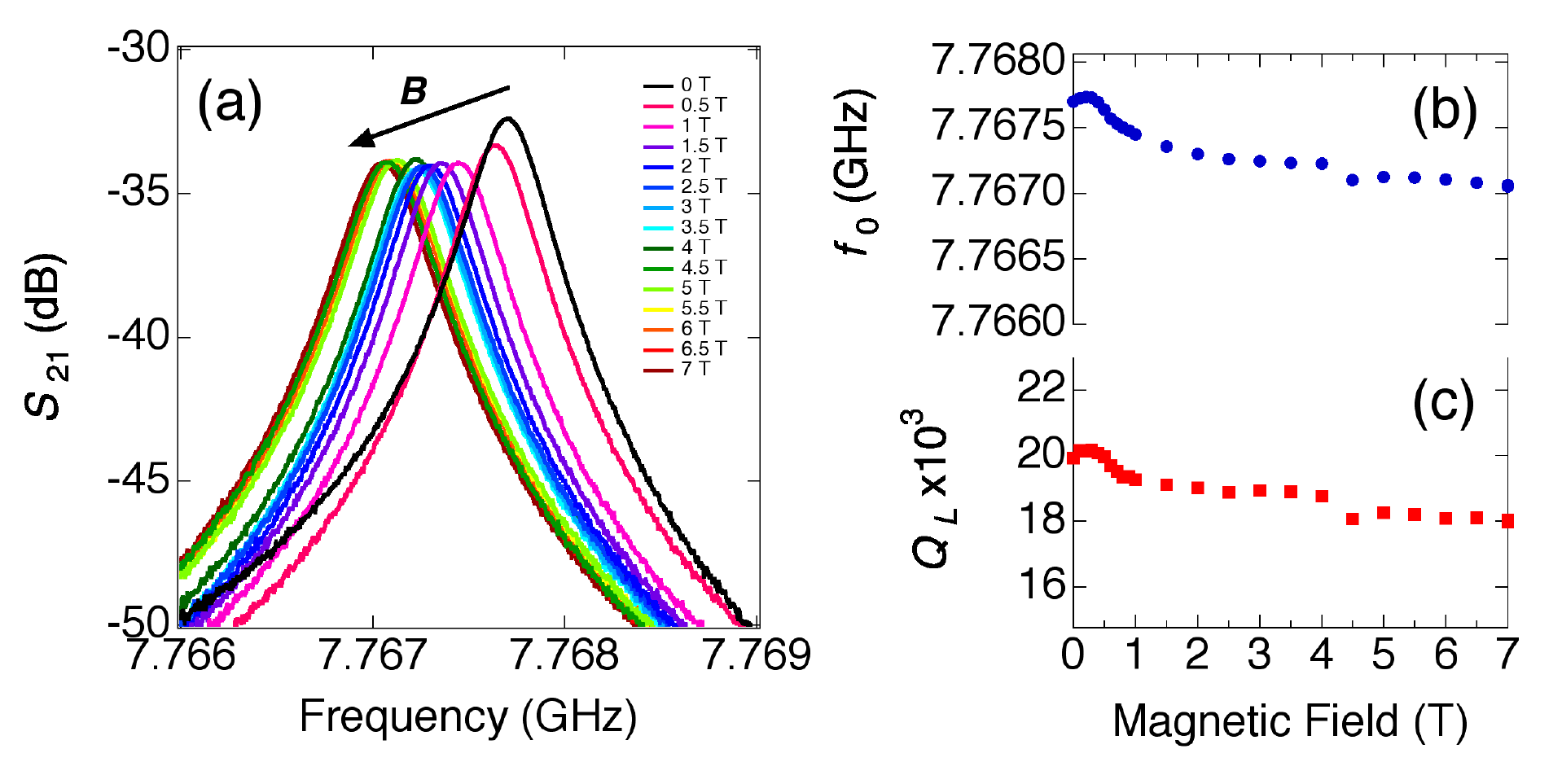}
\end{center}
\caption{(a) Transmission spectra as a function of frequency measured at 2 K for applied external magnetic fields up to 7 T. Dependence of (b) the cavity mode resonant frequency, $f_0$, and (c) the loaded quality factor, $Q_L$, on the externally applied magnetic field, $B$.}
\label{Fig2}
\end{figure}

We first show that coplanar resonators made of YBCO allow expanding temperature, magnetic field, and power ranges with respect to the Nb cavities commonly used in circuit-QED experiments. At $T=2$ K, the transmission spectrum shows a well defined resonance centered at $f_0$=7.7553 GHz [Fig.~\ref{Fig1}(a)]. The resonance dip is visible also in the $S_{11}(f)$ spectrum. This indicates that the resonator is not undercoupled, and that the loaded quality factor ($Q_L$) should be considered. Since $S_{21}(f)=20 \log_{10}( \sqrt{P(f)/P_{inc}})$, where $P(f)$ and $P_{inc}$ are the transmitted and the input powers at the capacitor of the resonator, respectively,\cite{SI} the transmission spectrum can be fitted by\cite{Hein}
\begin{equation}
S_{21}(f)=-IL-10 \log_{10}\left[1+Q_L^2\left(\frac{f}{f_0}-\frac{f_0}{f}\right)^2\right],
\label{lorentzian}
\end{equation}
where $Q_L$(2K)$\simeq 20000$ and the insertion loss is $IL=-S_{21}(f_0)=16.5$ dB [red line in Fig. \ref{Fig1} (b)]. The quality factor corresponds to that calculated from the half power bandwidth (3-dB method). More decoupled resonators display stable $Q_L>30000$ in the high power regime.\cite{SI}

The temperature dependence of the transmission resonance was investigated by measuring $S_{21}(f)$ in the range 2-100 K. To extract $f_0$ and $Q_L$ we fitted each spectrum with Eq. \ref{lorentzian}. Figure~\ref{Fig1}(b) shows a small shift of $f_0$ between 2 and 60 K. For higher temperature the resonance peak shifts towards lower frequencies, and it disappears in correspondence to the YBCO film critical temperature ($T_c=87$ K). The loaded quality factor [Fig.~\ref{Fig1}(c)] progressively decreases with increasing $T$, while remaining $Q_L(T)>10000$ for $T<55$ K. This behavior is in line with similar results reported in the literature.\cite{Lancaster} 

An applied magnetic field ($B$) generally gives rise both to a decrease of the quality factor and to a hysteretic behavior of the resonant frequency. While this behavior was effectively observed for intermediate temperatures,\cite{SI} at low temperature the field dependence becomes progressively weaker. Figure~\ref{Fig2}(a) shows a series of $S_{21}(f)$ spectra measured at 2 K for increasing $B$, up to 7 T. The values of $f_0$ and $Q_L$ extracted from Eq.~\ref{lorentzian} are plotted as a function of $B$ in (b) and (c): they are remarkably stable up to 7 T, being $Q_L(7 \mathrm{T}) = 0.90 \times Q_L(0 \mathrm{T})$. We notice that for Nb resonators a drop of $Q_L$  is observed for fields in the mT range.\cite{Bothner} 
Degradation of the quality factor of the superconducting resonators against the applied field is due to the dissipation mechanisms related to the vortex motion. This effect has been generally described in terms of increase of the surface resistivity ($R_s$) under applied magnetic field, whilst recent experimental results have evidenced that more sophisticated models are required for thin films.\cite{Krupka} Surface resistivity measurements performed at 20 K by means of the dielectric resonator method have shown a weak dependence of $R_s$ with respect to a dc field up to 5 T applied parallel to the YBCO film.\cite{Sato} These findings, independently obtained by different experimental techniques, corroborate the field dependence of $Q_L$ we report in Fig. \ref{Fig2}.

Summarizing the results of the YBCO resonators characterization, the decay rate of the cavity remains reasonably smaller than $\kappa/2\pi \sim 1$ MHz in a wide temperature and magnetic field ranges, and we can therefore exploit these unique properties to perform circuit-QED experiments with spin ensembles as follows. 

\begin{figure}[t]
\begin{center}
\includegraphics[width=8.5cm]{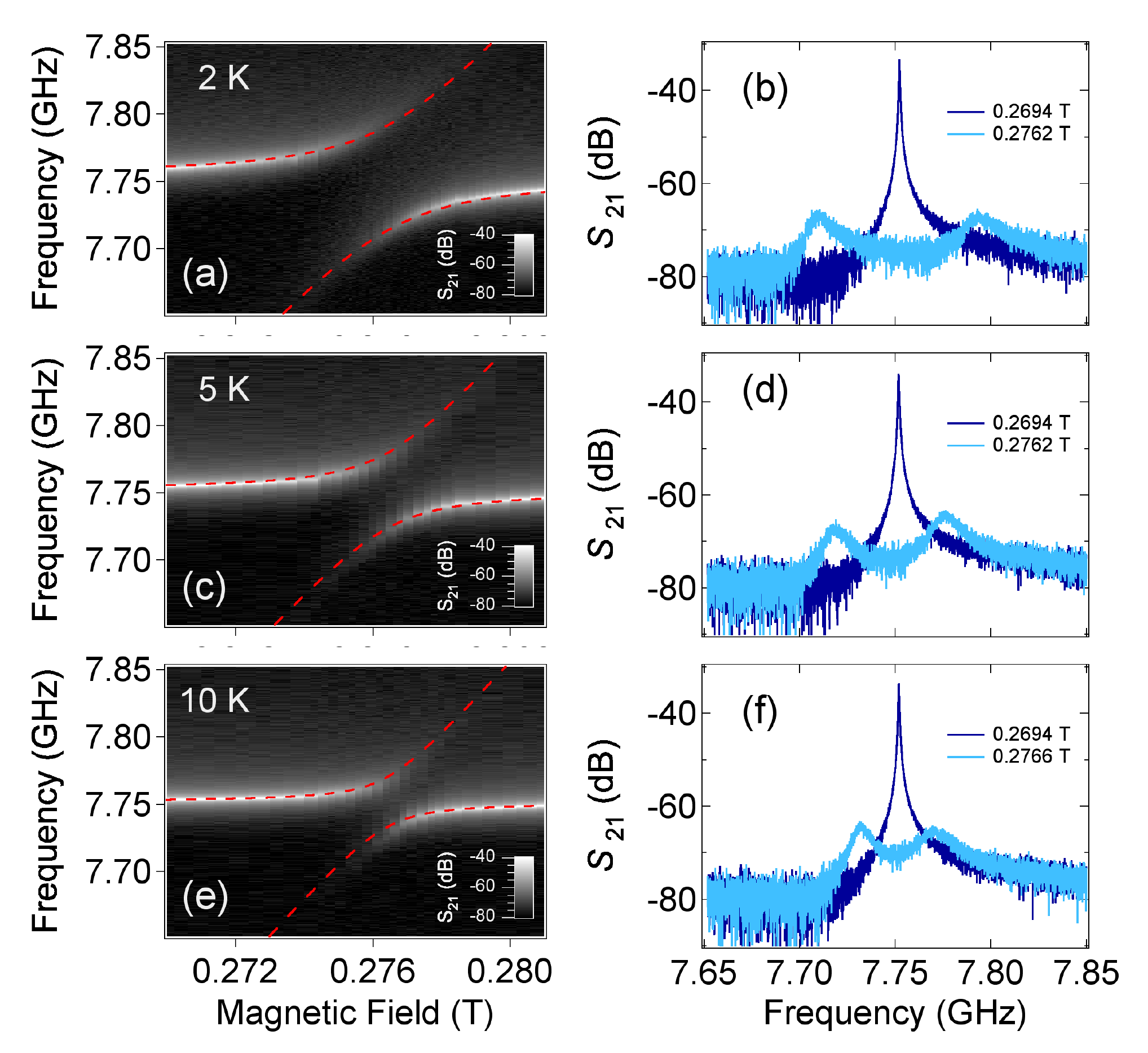}
\end{center}
\caption{Transmission spectra of a YBCO microwave resonator loaded with a DPPH spin ensemble ($P_{inc}=-12.5$ dBm). The left column shows a series of two-dimensional maps obtained by plotting $S_{21}(f)$ for different applied $B$. Temperature: 2 K (a), 5 K (c) and 10 K (e). Dashed red lines display the calculated curves. The right column shows the cross sections related to the corresponding right panel, either on resonance (cyan) or off resonance (blue).}
\label{Fig3}
\end{figure}

\begin{figure}[t]
\begin{center}
\includegraphics[width=8.5cm]{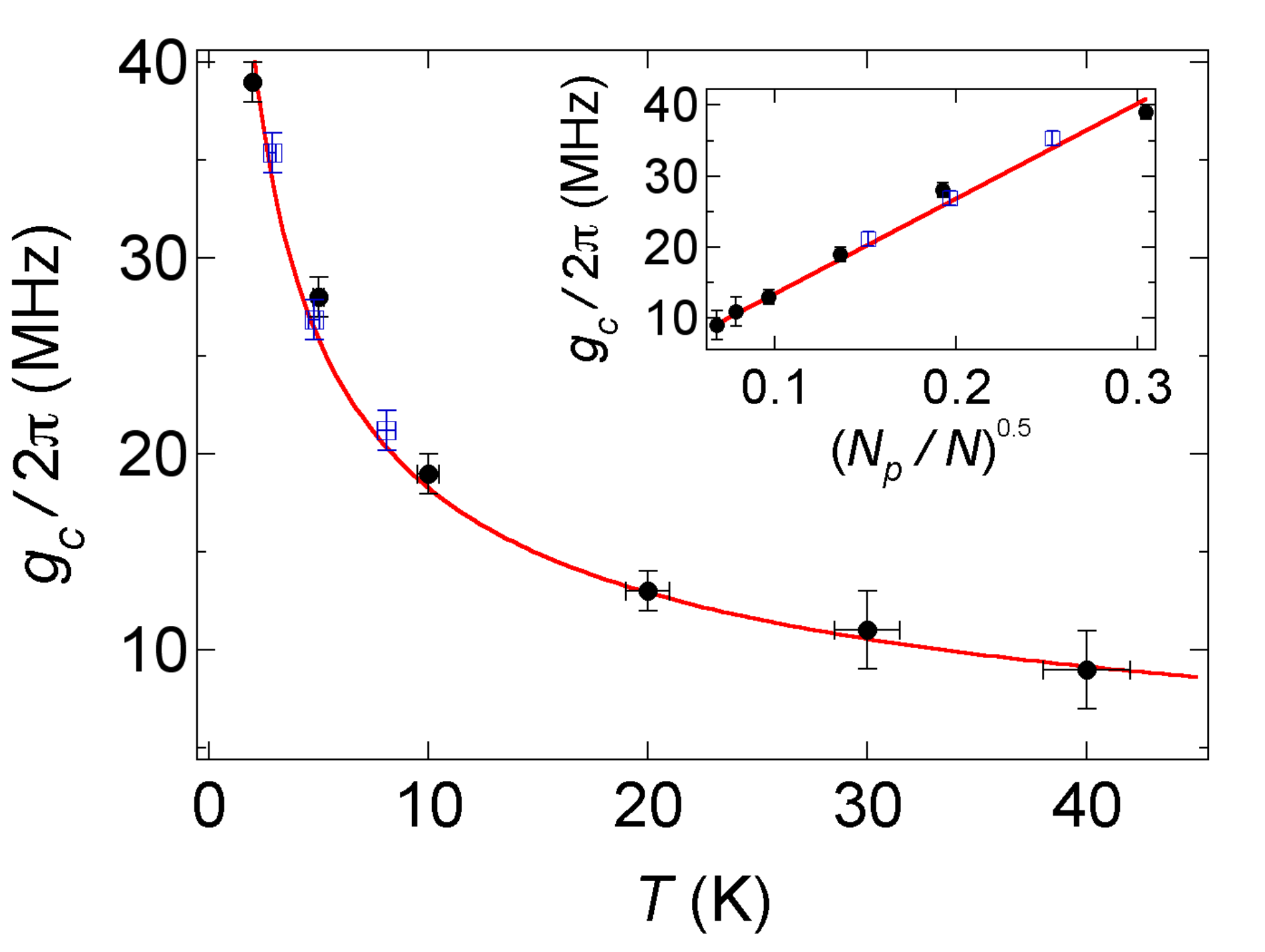}
\end{center}
\caption{Dependence of the collective coupling constant $g_c(T)$ (black circles) with respect to $T$ and $\sqrt{N_p(T)/N}$ (inset). The blue open squares display $g_c^*(T)/\sqrt{0.75}$. Solid red lines show the calculated curves.}
\label{Fig4}
\end{figure}

A thin layer of DPPH powder was attached to the center of the YBCO resonator by means of silicone grease. The volume of the sample ($1.2\times 0.5\times 0.05$ mm$^3$) was estimated under optical microscope, and it corresponds to a total number of approximately $N\simeq 6\times 10^{16}$ radicals.\cite{SI}
In Fig.~\ref{Fig3} we report the evolution of the transmission peak in correspondence to the resonance field of the DPPH spin ensemble ($B_r \simeq 0.276$ T). At 2 K two branches are observed, which indicate the presence of a large anticrossing between the resonator mode and the spin ensemble [Fig.~\ref{Fig3}(a)]. Cross-sectional $S_{21}(f)$ spectra measured on  resonance show two peaks separated by $f_{+} - f_{-} \simeq 82$ MHz [Fig.~\ref{Fig3}(b)].  The fit of the off-resonance (0.2694 T) spectrum with Eq. \ref{lorentzian} gives $f_0=7.7522$ GHz, $Q_L=16000$ and $IL=-33.5$ dB. Thus, the cavity decay rate results $\kappa\simeq 0.5$ MHz, while the estimated number of photons inside the cavity mode volume\cite{Sage} is $N_{ph}\simeq 10^{11}$. For increasing temperature, the width of the anticrossing decreases and the splitting is progressively reduced to $\simeq58$ MHz (at 5 K) and $\simeq 39$ MHz (at 10 K) [Fig. \ref{Fig3}(d) and (f)].  The splitting of the transmission peak in correspondence to the resonance field of DPPH is observed up to 50 K.\cite{SI} To evaluate the magnitude of the collective coupling constant $g_c$, we fitted the resonance frequency by using the usual expression for the vacuum field Rabi splitting (neglecting the damping rates, i.e. the imaginary parts of the split eigenfrequencies)\cite{Abe} 
\begin{equation}
\omega_{\pm}=\omega_0+\frac{\Delta}{2} \pm \frac{\sqrt{\Delta^2+ 4 g_c^2}}{2},
\label{coupling}
\end{equation}
where $\omega_{\pm}=2\pi f_{\pm}$, $\Delta=g \mu_B (B-B_r)/ \hbar$ and $g=2.0037$ is the  Land\'e $g$-factor of DPPH. The fitted rates $g_c / 2\pi$ are plotted as a function of temperature in Fig.~\ref{Fig4} (black circles), spanning from 39 MHz at 2 K to 9 MHz at 40 K. Hence, the strong coupling condition,\cite{RevMPnori}  $g_c \gg \gamma_s, \kappa$, is clearly satisfied in our sample in the whole temperature range up to 40 K.

To interpret the results in Fig.~\ref{Fig4}, we notice that the number of polarized $s=1/2$ spins ($N_p$) varies with temperature as 
\begin{equation}
\frac{N_p}{N}=\tanh\left(\frac{h f}{2 k_B T}\right).
\label{tanh}
\end{equation} 
Simultaneously, $g_c$ rescales non-linearly as 
\begin{equation}
g_c=g_s \sqrt{N_p},
\label{collective_g}
\end{equation} 
where $g_s$ is the coupling of a single spin 1/2 to the resonator mode. In Fig. \ref{Fig4} we calculated the behavior of $g_c$ by means of Eqs. \ref{tanh} and \ref{collective_g} (solid red line). The best fit was obtained by means $g_c/2\pi$[MHz]$=134\sqrt{N_p/N}$. From the zero temperature limit ($N_p = N\sim 6\times 10^{16}$), and the finite temperature values, we estimated a single-spin coupling rate in a restricted range $g_s /2\pi \sim 0.55-0.60$ Hz. These values well compare with the one independently estimated from the known parameters of our resonators,\cite{SI} besides being consistent with the values reported in the literature for different cavity geometries and spin systems.\cite{RevMPnori} We also notice that even at the highest temperatures at which strong coupling can be observed, when only a small fraction of spins is coupled to the microwave mode, the condition $N_{ph}\ll N_p$ still holds, for which Eq.~\ref{collective_g} is a safe assumption.\cite{Chiorescu}

To get further insight about the $g_c(N_p)$ dependence, we removed a portion of DPPH sample from the resonator and we repeated the measurements to extract the reduced coupling constant $g_c^*(T)$.\cite{SI} The remaining sample corresponds to approximatively $75\%$ of the original volume. The total number of $s=1/2$ radicals is thus reduced to $N^*=0.75 \times N$ and  $g_c^*=g_s\sqrt{0.75\times N_p}$. To compare these results with  $g_c(T)$, in Fig.~\ref{Fig4} we plotted the rescaled coupling constant $g_c^*/\sqrt{0.75}$ as function of $T$ and $\sqrt{N_p/N}$ (open squares). The very good agreement between $g_c^*(T)$ and $g_c(T)$ corroborates the behavior described from Eqs. \ref{tanh} and \ref{collective_g}.

In summary, we have shown that YBCO microwave planar resonators constitute a viable route for the implementation of on-chip quantum technologies. In particular, we have shown their robustness against an external magnetic field up to 7 T and up to liquid Nitrogen temperature, which makes them ideal candidates for circuit-QED experiments with spin ensembles. To display their potentialities, we showed that the collective strong coupling regime of a DPPH ensemble coupled to coplanar YBCO resonators can be achieved up to 40 K.

\textbf{Acknowledgements. }
The authors warmly acknowledge S. Carretta, A. Chiesa, A. Lascialfari, F. Troiani, and M. Barra for useful discussions, T. Orlando for supplementary EPR measurements and S. Marrazzo for fabrication expertise.
This work was funded by the Italian Ministry of Education and Research (MIUR) through ``Fondo Investimenti per la Ricerca di Base'' (FIRB) project RBFR12RPD1, and by the US AFOSR/AOARD program, contract FA2386-13-1-4029.

\newpage\null\thispagestyle{empty}\newpage

\onecolumngrid

\part*{Supplementary Information}

\section{Experimental section}

Experiments with YBCO resonators and applied magnetic fields up to 7 T were carried out by means of the low temperature set-up shown in Fig. S1. The external magnetic field is applied parallel to the YBCO film, and used to tune the DPPH spin gap. The coplanar resonators were installed into an oxygen free high conductivity copper box that allows installation, grounding, as well as thermalization down to 2 K. The resonator box is installed on a dedicated probe wired with cryogenic coaxial cables (Micro-Coax UT-085B-SS), and inserted in a Quantum Design PPMS cryo-magnetic set-up. The temperature is monitored by means of a RuO$_2$ thermometer located close to the resonator box. Reflection ($S_{11}$) and transmission ($S_{21}$) scattering parameters were measured by means of a Vector Network Analyzer (Agilent PNA 26.5 GHz). The microwave launchers are constituted by a SMA connector on the coaxial cable side, and a gold-plated pin on the device side. The relative length of the pin can be regulated to tune the effective coupling between resonator and feedline. 

To calibrate the transmission spectrum and to remove the insertion loss of the coaxial line ($IL_{coax}$), we measured the $S_{21}(f)$ spectrum of a calibration device constituted by a superconducting waveguide without capacitive gaps, which was installed in place of the coplanar resonator. The incident power at the input capacitor of the resonator was estimated as $P_{inc}=P_{out}-IL_{coax}/2$, where $P_{out}$ is the output power of the Vector Network Analyzer.

\begin{figure}[pbb]
\begin{center}
\includegraphics[width=8cm]{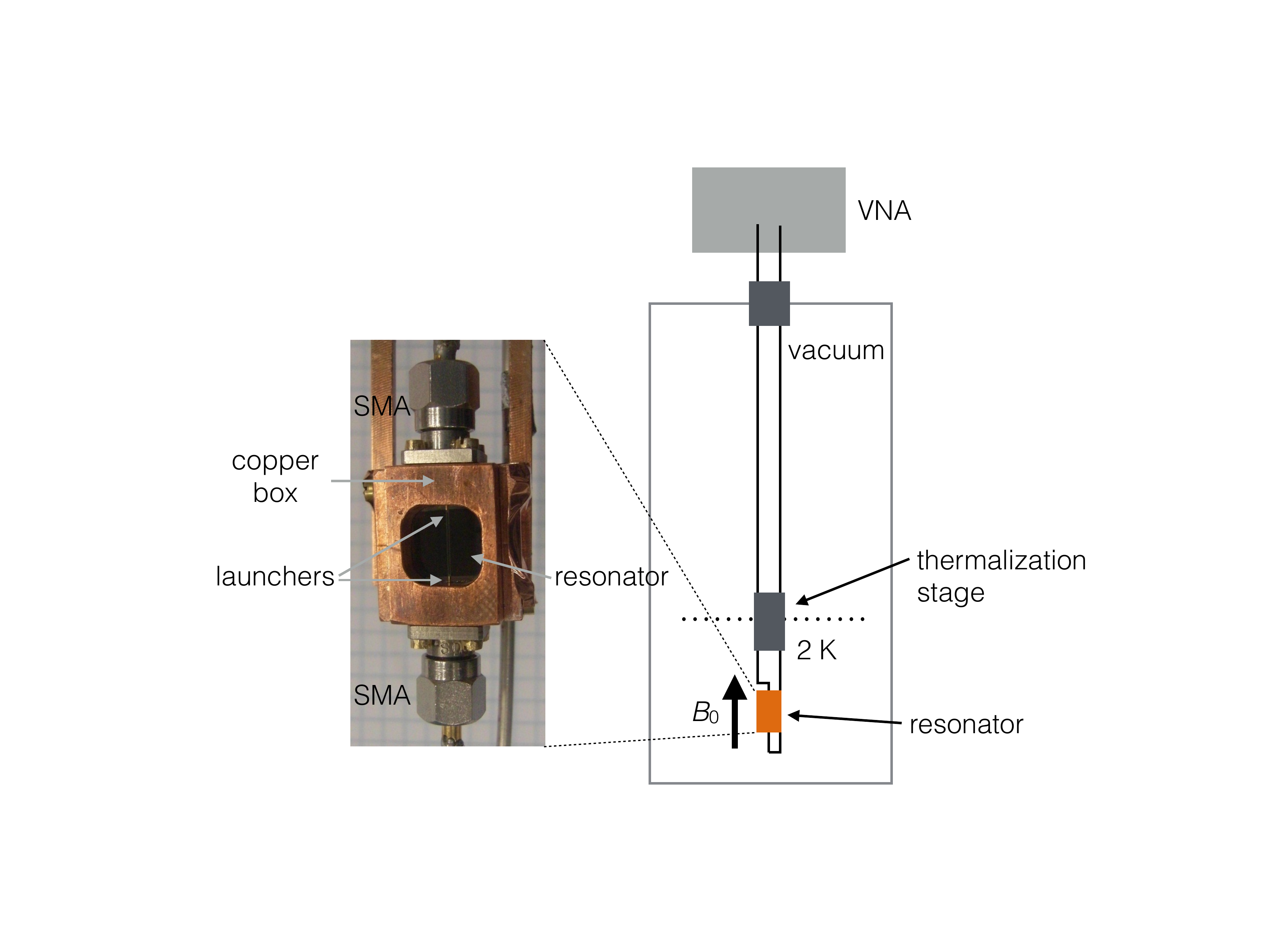}
\end{center}
FIG. S1. Left. Photograph of the YBCO coplanar resonator installed into the copper box. Right. Schematic of the experimental set-up. The base temperature is 2 K and the maximum field strength is 7 T.
\label{schematic}
\end{figure}

\section{Characterization of the YBCO coplanar resonators}

We fabricated and tested five YBCO/sapphire coplanar resonators (named resonator \#1 to \#5) with the geometric dimensions reported in the letter (the resonator \#1 has been tested for two different coupling configurations, \#1a and \#1b). All the resonators show comparable behavior in terms of dependence of resonance frequency ($f_0$) and quality factor ($Q_L$) from temperature ($T$) and applied magnetic field ($B$). The strong coupling between microwave field and DPPH radicals has been reproducibly obtained with all the tested resonators.

The measured values of resonant frequency $f_0$ and loaded quality factor $Q_L$ are shown in Fig. S2 for resonator \#4. In the range 2-10 K both $f_0$ and $Q_L$ show a weak dependence from $T$. Fig. S3 shows the dependence of the loaded quality factor measured as a function of the applied magnetic field. These experimental data have been measured on resonator \#1 with two different coupling capacitances (\#1a and \#1b). A small hysteresis behavior of $f_0$ has been observed by repetitively cycling the magnetic field between +0.5 and -0.5 T.  Its shape qualitatively resembles the hysteresis reported for Nb resonators \cite{BothnerPRB}. For the YBCO/sapphire resonator, the maximum variation of $f_0(B)$ is very small (0.44 MHz), and it corresponds to 0.006$\%$ of the resonance frequency at zero field. The degradation of the internal quality factor of high temperature superconductors under applied magnetic field has been widely investigated and different models have been proposed \cite{Coffey, Lancaster, Vendik, Blatter, Ryu}. The generalized two fluid model, which includes quasi-particle excitations and vortex motion, has been often used to interpret the microwave surface resistance of YBCO superconducting thin films under high dc magnetic field \cite{Onshima, Honma, Sato}. The complex conductivity approach has been shown to better describe the properties of YBCO films under applied magnetic field \cite{KrupkaIEEE, KrupkaAPL}. Magnetic flux penetrates in the superconductor through vortices, with the size of about few nanometers in YBCO \cite{Plourde, Wells}; their motion produces dissipation resulting in an increase of the surface resistance (flux flow). In a real system defects are always present and cause the vortices to become pinned at particular positions. In addition, in layered structures such as the high-$T_c$ superconductors the vortex can be pinned by the block interspacing between the superconducting CuO$_2$ layers when the field is perpendicularly applied. However, vortex lines can move between these pinning centers with thermally activated jumps (flux creep process). This is expected to be the main process causing both the large reduction of the quality factor as the temperature is increased, as well as the sensitive degradation as a function of the applied field at high temperature, as displayed in Fig. S3. Quantum tunneling may also cause flux motion at very low temperature (quantum creep process) \cite{Hoekstra}. The occurrence of these effects has a strong dependence on temperature and applied magnetic field, as well as on the characteristics of the superconducting material. 

\begin{figure}[ptb]
\begin{center}
\includegraphics[width=8cm]{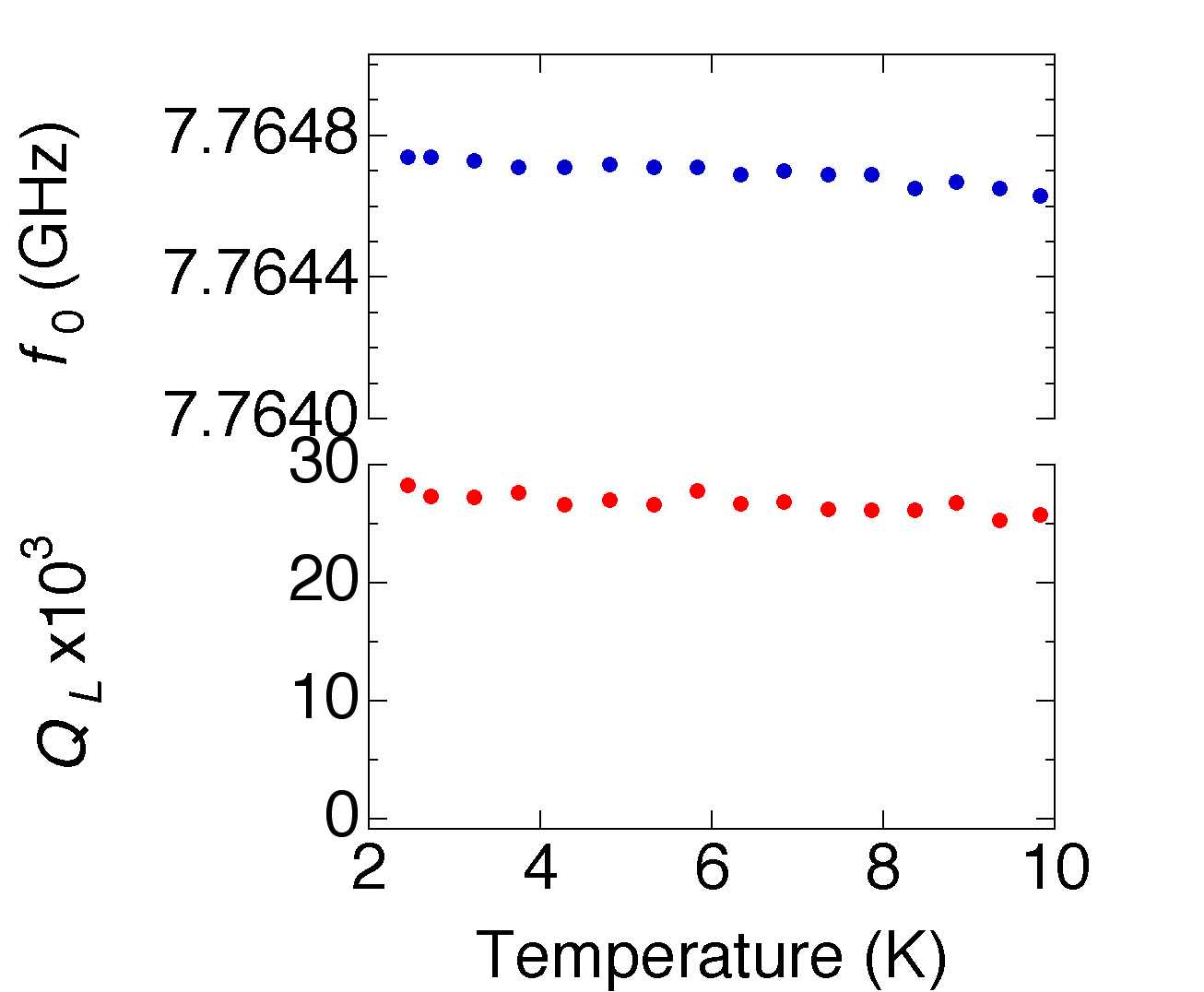}
\end{center}
FIG. S2. Temperature dependence of resonant frequency (upper panel) and loaded quality factor (lower panel) of resonator \#4.
\label{res4_T}
\end{figure}

\begin{figure}[pbb]
\begin{center}
\includegraphics[width=8cm]{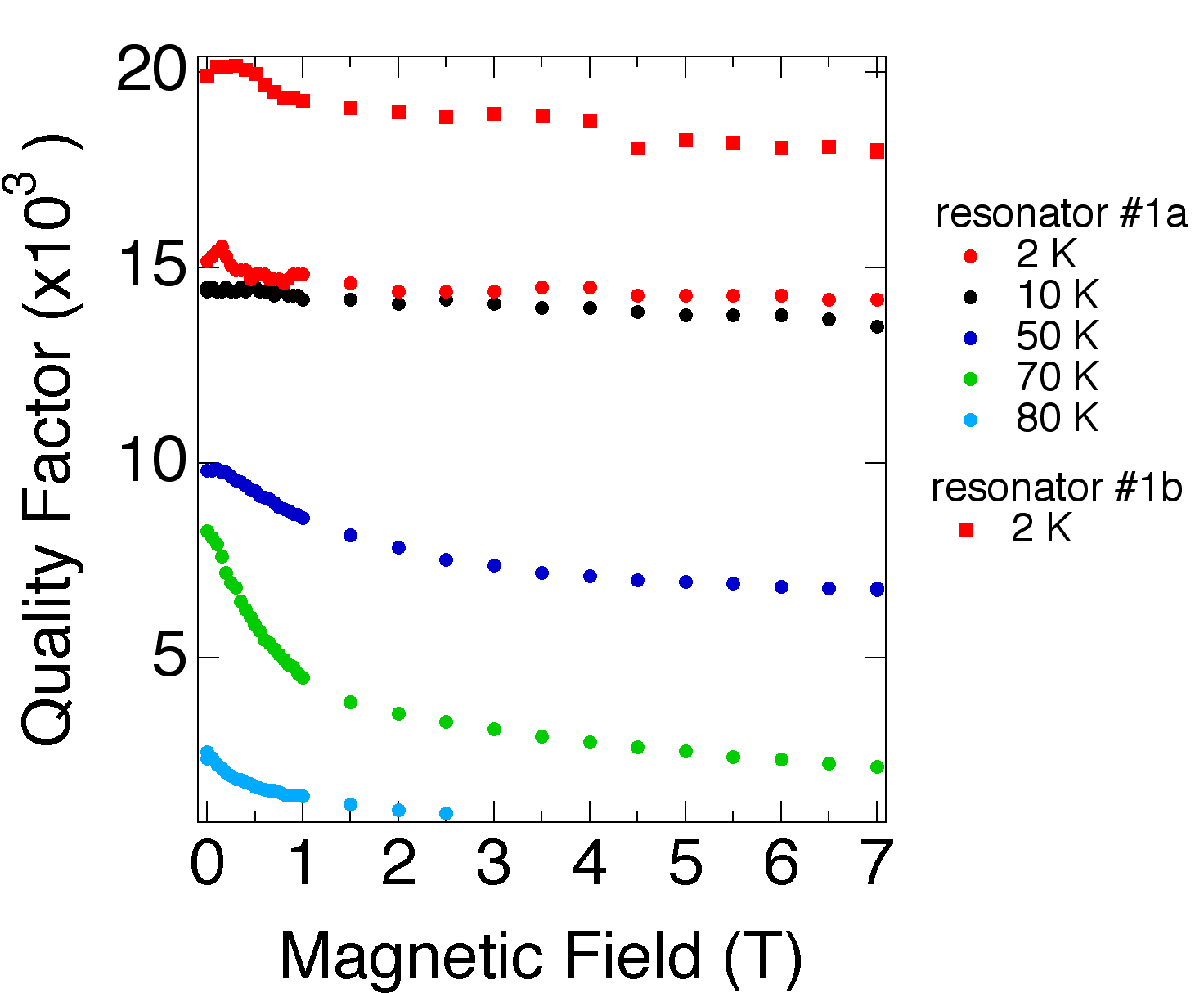}
\end{center}
FIG. S3. $Q_L$-vs-$B$ curves measured for different temperatures on the bare resonator \#1a (circles) and \#1b (squares)
\label{Q_field_temp}
\end{figure}

\begin{figure}[pbb]
\begin{center}
\includegraphics[width=14cm]{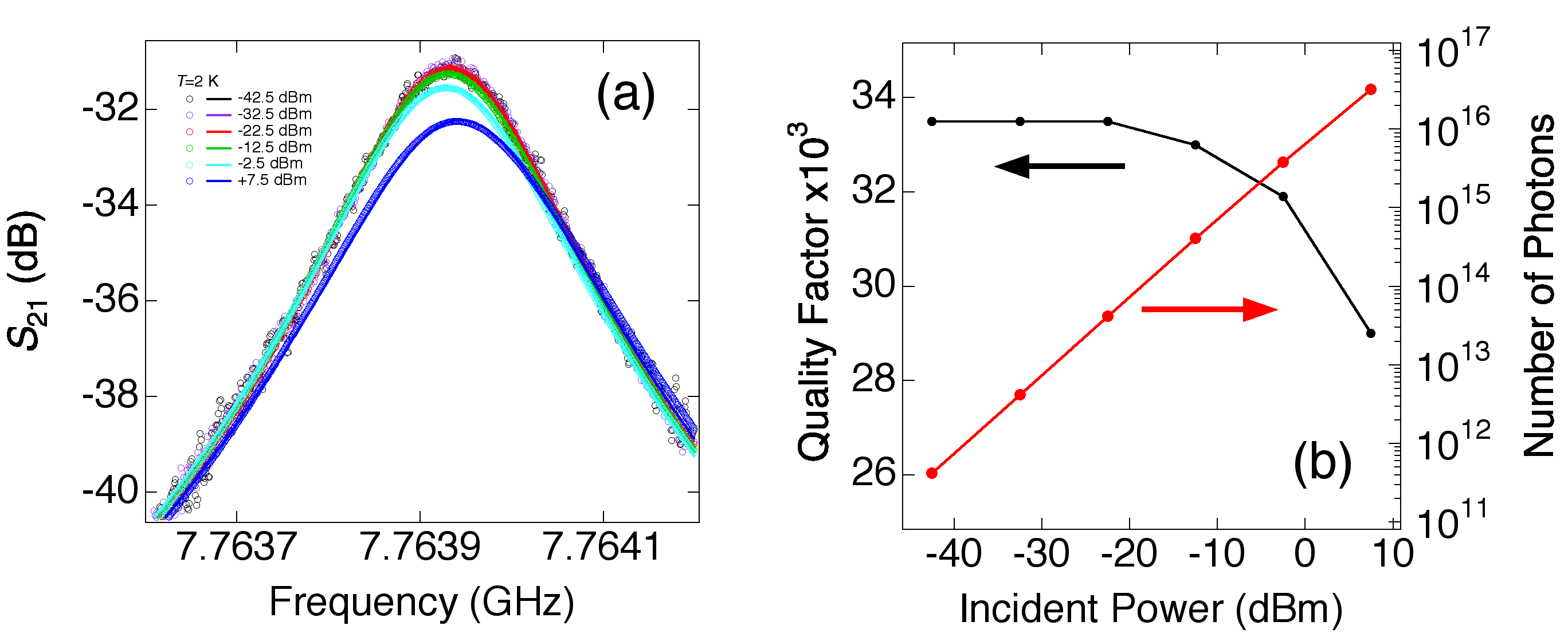}
\end{center}
FIG. S4. (a) $S_{21}(f)$ spectra measured for resonator \#5 with several values of $P_{inc}$ (open circles). The temperature is 2 K. Solid lines show the fit curve calculated from Eq. 1. (b) dependence of the fitted $Q_L$ and of the number of photons calculated from Eq. 2.
\label{power}
\end{figure}

\section{Estimation of the number of photons}

Figure S4 (a) shows the dependence from the incident power of the transmission spectrum of resonator \#5. From the fit of the transmission spectra with 
\begin{equation}
S_{21}(f)=-IL-10 \log_{10}\left[1+Q_L^2\left(\frac{f}{f_0}-\frac{f_0}{f}\right)^2\right],
\label{lorentzian}
\tag{S1}
\end{equation}
we extracted $f_0=7.76393$ GHz and $IL=31.3$ dB. The fitted valued of the quality factor is $Q_L=33500$ for $P_{inc}<-22.5$ dBm, while for very high power $Q_L$ decreases [Fig. S4 (b)]. The number of photons 
\begin{equation}
N_{ph}=\frac{P_{circ}}{h f_0^2}
\tag{S2}
\end{equation}
can be calculated from the circulating power \cite{Sage}
\begin{equation}
P_{circ}=\frac{1}{\pi} P_{inc} Q_L 10^{-IL/20}, 
\tag{S3}
\end{equation}
and results comprised in the range $10^{11}$-$10^{17}$ respectively for $P_{inc}$ varying between -42.5 and 7.5 dBm  [Fig. S4 (b)].

\section{Estimation of the number of radicals}

The total number of spins in our sample can be estimated from the known radicals density in DPPH, $\rho= 1.4$ g/cm$^3$ \cite{Kiers}. Knowing the molar mass $m_{mol}=394.32$ g/mol, we get the density as $\rho_V \sim 2\times 10^{21}$ cm$^{-3}$. Then, the number of radicals in a sample of known volume, $V$, is simply $N=\rho_V V$.

\section{Linewidth of DPPH organic radical}

Spin-phonon relaxation ($T_1$) and dephasing ($T_2$) times of DPPH organic radicals are reported in the literature. For concentrated samples $T_1=T_2=62$ ns \cite{EatonEaton}. By considering the homogeneous broadening, the linewidth can be calculated from \cite{WeilBolton}
\begin{equation}
\gamma_s=\frac{1}{2T_1}+\frac{1}{T_2} \sqrt{1+\gamma^2 B_1^2 T_1 T_2}
\tag{S4}
\label{broadening}
\end{equation}
where $\gamma=g\mu_B/\hbar$ is the gyromagnetic ratio and $B_1$ is the amplitude of the microwave field. Due to the effect of the exchange narrowing \cite{Anderson}, the calculated linewidith is overestimated respect to that obtained experimentally. Recent measurements report $\gamma_s=3.9$ MHz in the range 10-295 K \cite{Zilic}. For $T<10$ broadening of the EPR line occurs due to antiferromagnetic interactions between the DPPH radicals. The reported linewidth is $\gamma_s=14$ MHz at 2 K \cite{SovPhysJEPT}.

\section{Estimation of the spin-photon coupling rate}

The single spin-photon coupling rate in the resonator, $g_s$, can be assumed to be equal to half the Rabi frequency of a spin 1/2 under the resonator vacuum magnetic field\cite{Tosi}
\begin{equation}
g_s=\frac{g\mu_B B_{vac}}{4h} \, ,
\label{g_s}
\tag{S5}
\end{equation}
where $g=2.0037$ is the DPPH single radical gyromagnetic factor, $\mu_B=9.27 10^{-24}$ J/T the Bohr magneton, and $h=6.62 10^{-34}$ J s the Planck constant. An approximate expression to evaluate $B_{vac}$ from the resonator parameters can be given as \cite{Tosi} 
\begin{equation}
B_{vac}\simeq\frac{\mu_0 I_{vac}}{2w} \, ,
\label{B_vac}
\tag{S6}
\end{equation}
where we $\mu_0=4 \pi \times 10^{-7}$ N A$^{-2}$ is the vacuum magnetic permeability, $I_{vac}=8.8\times 10^{-8}$ is the zero-point current in the resonator, which can be estimated as $I_{vac}=\pi \sqrt{h/Z_0} f_0$, $Z_0$ being the characteristic impedance of the transmission line (see also Ref.~\onlinecite{Tosi}), and we assumed $w=400$ $\mu$m (taking into account the transverse extension of the $B$-field lines on the coplanar resonator surface, including the central strip line of 200 $\mu$m, the gaps and part of the ground planes). From these values we obtain $B_{vac}\simeq 1.35 \times 10^{-10}$ T. 
Hence, the single spin coupling rate results from Eq. S5 $g_s\simeq 0.9$ Hz. This value is remarkably close to the 0.6 Hz estimated from the experimental collective coupling rates, $g_c$, considering that the simple analysis above neglects any average on the spatial distribution of the spins within the magnetic field profile.

\begin{figure}[ptb]
\begin{center}
\includegraphics[width=7.8cm]{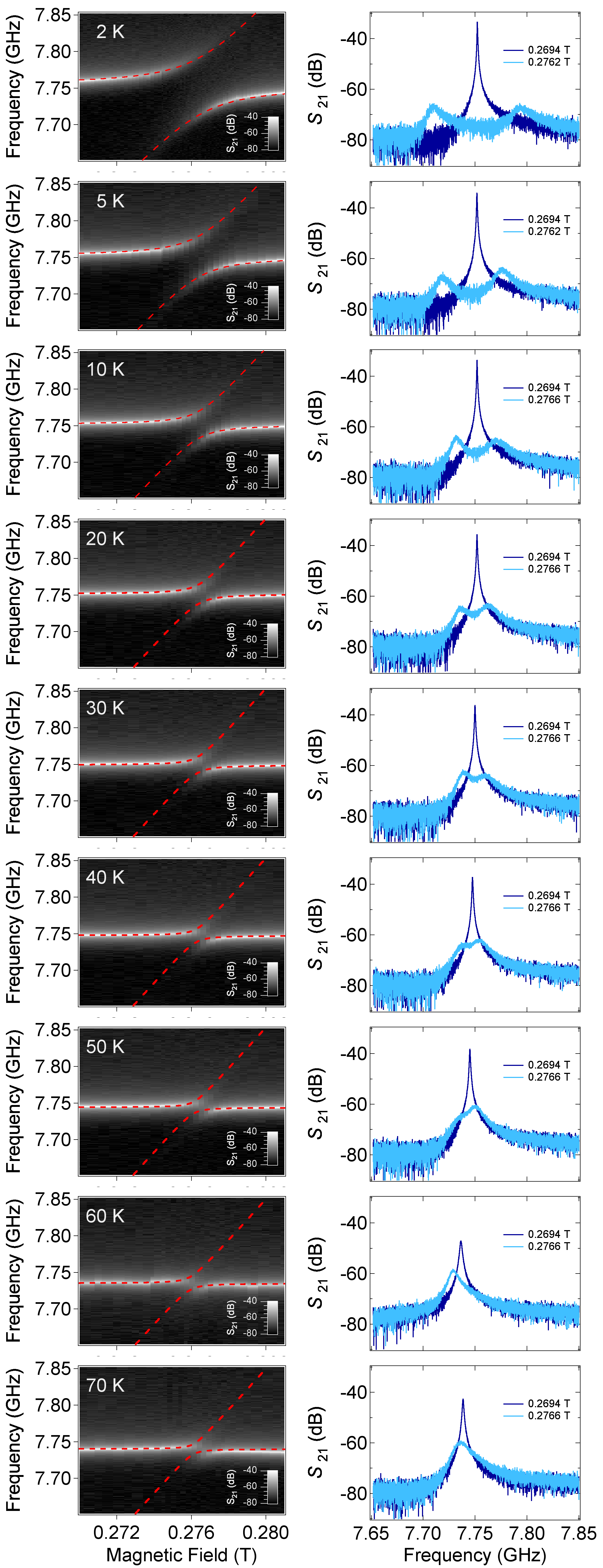}
\end{center}
FIG. S5. Experimental data obtained on the resonator \#2 loaded with the DPPH powder sample (ensemble of $N$ spins). For each temperature (rows), the panel in the left column shows the grey scale plot of the $S_{21}$-vs-$f$ spectra taken at different $B$, whilst the correspondent right panel displays the cross sectional spectrum. The splitted $S_{21}$ spectrum was observed in the temperature range 2-40 K in correspondence to the resonance field of DPPH. At $T=50$ K the peaks coalesce and for higher temperature a single peak is observed also at resonance.
\label{FigS5}
\end{figure}
\begin{figure}[ptb]
\begin{center}
\includegraphics[width=8.5cm]{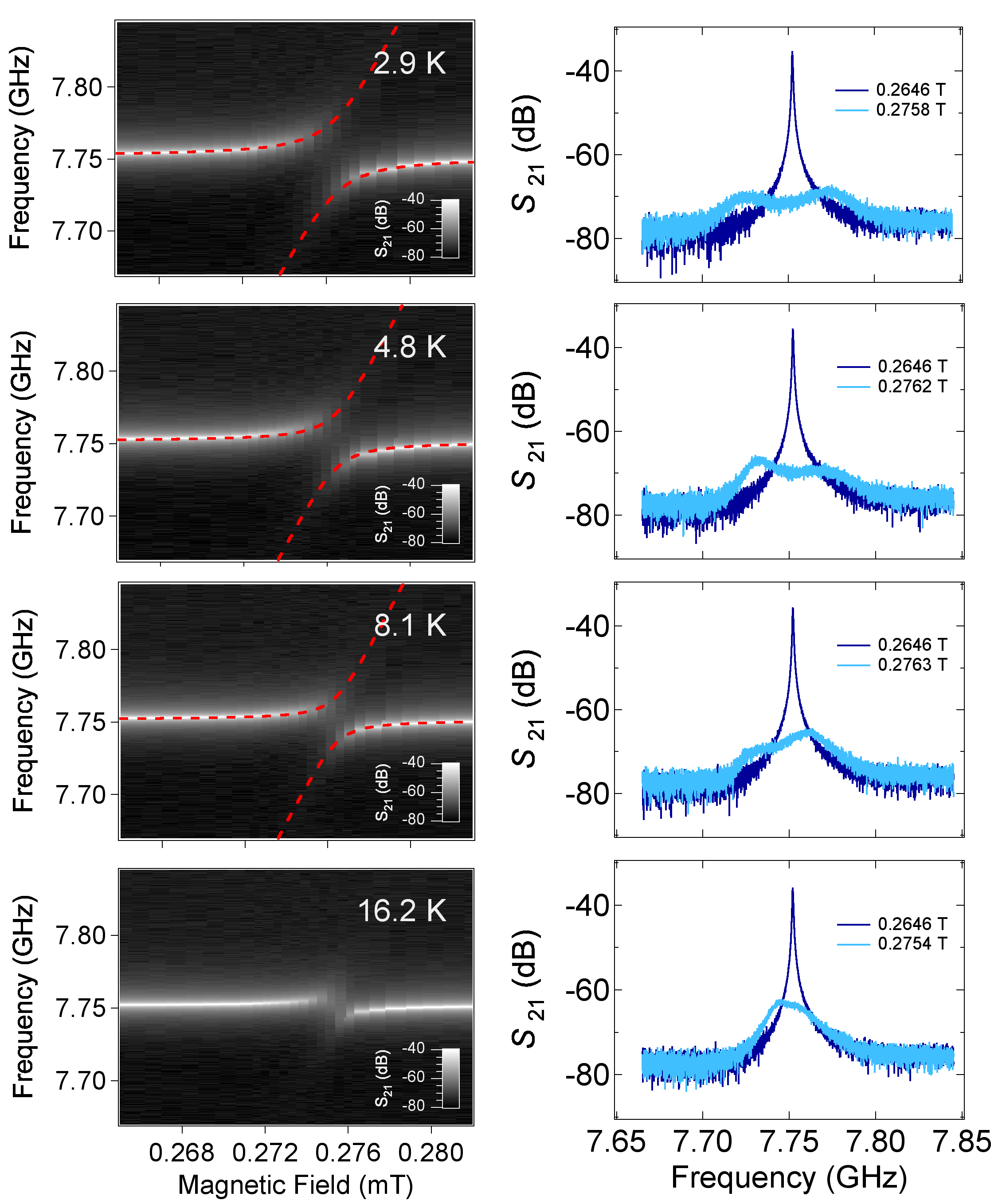}
\end{center}
FIG. S6. Experimental data obtained on the resonator \#2 loaded with the reduced DPPH powder sample (ensemble of $N^*=0.75 \times N$ spins).  For each temperature (rows), the panel in the left column shows the grey scale plot of the $S_{21}$-vs-$f$ spectra taken at different $B$, whilst the correspondent right panel displays the cross sectional spectrum measured on and off resonance. Red dashed lines shows the best fit obtained to extract $g_c^* (T)$ by means of Eq. 2 of the letter.
\label{FigS6}
\end{figure}

\end{document}